\documentclass[useAMS,usenatbib,usegraphicx]{mn2e}

\usepackage{amssymb}
\usepackage{amsmath}
\usepackage{bm}
\newcommand{\FAP}{{\rm FAP}}

\title[Multi-frequency detection significance]{Detecting multiple periodicities in
observational data with the multi-frequency periodogram.\\ I. Analytic assessment of the
statistical significance}

\author[R.V.~Baluev]{Roman V. Baluev\thanks{E-mail: roman@astro.spbu.ru}\\
Central Astronomical Observatory at Pulkovo of Russian Academy of Sciences, Pulkovskoje
shosse 65, St Petersburg 196140, Russia\\
Sobolev Astronomical Institute, St Petersburg State University, Universitetskij prospekt
28, Petrodvorets, St Petersburg 198504, Russia}

\begin{document}

\date{Accepted 2013 August 23.
      Received 2013 July 23;
      in original form 2013 May 20}

\pagerange{\pageref{firstpage}--\pageref{lastpage}} \pubyear{2013}

\maketitle

\label{firstpage}

\begin{abstract}
We consider the ``multi-frequency'' periodogram, in which the putative signal is modelled
as a sum of two or more sinusoidal harmonics with idependent frequencies. It is useful in
the cases when the data may contain several periodic components, especially when their
interaction with each other and with the data sampling patterns might produce misleading
results.

Although the multi-frequency statistic itself was already constructed, e.g. by G.~Foster
in his CLEANest algorithm, its probabilistic properties (the detection significance
levels) are still poorly known and much of what is deemed known is unrigourous. These
detection levels are nonetheless important for the data analysis. We argue that to prove
the simultaneous existence of all $n$ components revealed in a multi-periodic variation,
it is mandatory to apply at least $2^n-1$ significance tests, among which the most
involves various multi-frequency statistics, and only $n$ tests are single-frequency ones.

The main result of the paper is an analytic estimation of the statistical significance of
the frequency tuples that the multi-frequency periodogram can reveal. Using the theory of
extreme values of random fields (the generalized Rice method), we find a handy
approximation to the relevant false alarm probability. For the double-frequency
periodogram this approximation is given by an elementary formula $\frac{\pi}{16} W^2
e^{-z} z^2$, where $W$ stands for a normalized width of the settled frequency range, and
$z$ is the observed periodogram maximum. We carried out intensive Monte Carlo simulations
to show that the practical quality of this approximation is satisfactory. A similar
analytic expression for the general multi-frequency periodogram is also given in the
paper, though with a smaller amount of numerical verification.
\end{abstract}


\section{Introduction}
The raw time-series data that are obtained in astronomical or other observations often
contain more than one periodic component. For example, in the field of the exoplanets
discovery, the fraction of known multi-planet systems is close to $20\%$, with about $1.3$
planets per system in average (See \emph{The Extrasolar Planets Encyclopaedia} at
www.exoplanet.eu). Although some may claim that $20\%$ is still only a minor fraction, the
multi-planet systems are the objects which are most interesting for further investigations
and most important for the associated theory work.

However, the complicated character of the compound radial velocity variation that multiple
planets induce on their host star, sometimes makes such data rather difficult for an
analysis. It is well-known that multiple periodic variations can interfere with each other
and with the periodic patterns of the non-uniform time series sampling. When we plot a
traditional periodogram of such data, we may even discover that its maximum peak is
unrelated to \emph{any} of the real periodicities (even for data without any random noise
at all). Such examples are given, for example, in \citep{Foster95} and also in the
teaching manual \citep{Vit-nun}. This obviously appears because the single-sinusoid signal
model, that is implicitly used by the \citep{Lomb76}--\citep{Scargle82} periodogram, as
well as by its more advanced relatives, is inadequate when dealing with the data
containing two or more periodicities. Note that we do not speak here of the overtone
harmonics that appear when a single \emph{non-sinusoidal} periodic signal is involved; for
that case the so-called multi-harmonic periodogram \citep{SchwCzerny96,Baluev09a} should
be used. We address here a more general case when the frequencies of the harmonics are
independent and unknown (not binded to any basic frequency).

On itself, the definition of a periodogram that could take into account at least two or
many periodicities at once is rather trivial (see Section~\ref{sec_def}). We do not claim
any invention rights on it: the earliest, to our concern, papers utilizing essentially the
same multi-frequency statistic were \citep{Foster95,Foster96a,Foster96b}. In these works,
the multi-frequency test statistic is treated as a part of the Foster's CLEANest
algorithm. In fact, the ``global'' version of the CLEANest method implies the direct
dealing with the multi-frequency periodogram plotted in a multi-frequency space (or at
least in some representative portions of this space).

The non-trivial problem that we address here is the estimation of the false alarm
probability ($\FAP$) associated with the observed peaks of this periodogram. The $\FAP$ is
necessary to justify any claims of signal detection in the presence of the noise. For the
multi-frequency periodogram the $\FAP$ characterizes the joint significance of an
extracted group of periodicities, and it is different from the usual significances of
individual periodicities, treated each on its own. We discuss this difference and the
practical importance of the joint $\FAP$ in Section~\ref{sec_mot}.

Since the $\FAP$ is related to the distribution function of the periodogram maxima, we
estimate it with the generalized Rice method for random processes and fields, which allows
to efficiently approximate to the relevant extreme-value distributions. We first used this
method for the Lomb-Scargle periodogram in \citep{Baluev08a}, and after that it appeared
very useful. In Section~\ref{sec_conf} we derive the $\FAP$ formula for the
multi-frequency periodograms using our generalized method described in \citep{Baluev13b}.

In Section~\ref{sec_gen} we discuss some further generalizations of the basic
multi-frequency periodogram, which are analogous to the ones already known for the
Lomb-Scargle one. After that, in Section~\ref{sec_simul} we present the results of Monte
Carlo simulations that we carry out to verify our analytic $\FAP$ estimations, while in
Section~\ref{sec_sens} we compare the detection sensitivity of the single- and
multi-frequency periodograms. Finally, in Section~\ref{sec_examples} we provide a
demonstrative example showing the potential capabilities of the double-frequency
periodogram.

\section{Multi-frequency periodogram}
\label{sec_def}
First of all, we should define the dataset that we deal with. Let it consists of $N$
measurements $x_i$, taken at the time $t_i$, and having the uncertainty $\sigma_i$. For
now we assume that $\sigma_i$ are known accurately, although later we will also consider
the more wide-spread case when only the weights $w_i\propto 1/\sigma_i^2$ are known, while
the values of $\sigma_i$ are known only to a free multiplicative constant. The errors of
$x_i$ are assumed Gaussian and mutually independent, with the variances given by
$\sigma_i^2$.

When deriving the definition of the classic Lomb-Scarge periodogram, it is assumed that
the data contain only pure noise (the null hypothesis) or the noise as well as a signal
(the alternative hypothesis). The classic Lomb-Scargle periodogram is based on a simple
sinusoidal model of the signal to detect:
\begin{equation}
\mu = A \cos(2\pi f t) + B \sin(2\pi f t),
\label{sf_mod}
\end{equation}
where the parameters $A$ and $B$ are implicitly estimated by means of a linear regression.
Now let us assume that the compound signal, that we expect to reveal in the time-series
data, is representable as the following sum of $n$ independent sinusoids:
\begin{equation}
\mu = \sum_{i=1}^n A_i \cos(2\pi f_i t) + B_i \sin(2\pi f_i t).
\label{mf_mod}
\end{equation}
The definition of the multi-frequency periodogram itself, based on~(\ref{mf_mod}), is
analogous to the one for the Lomb-Scargle periodogram, based on~(\ref{sf_mod}). Here we
slightly extend the formulae of a general (but still single-frequency) linear periodogram,
given in \citep{Baluev08a}. Now we have $2n$ unknown linear coefficients $A_i$ and $B_i$.
In fact, the frequencies $f_i$ are unknowns too, but we treat them separately since they
are non-linear parameters. Let us rewrite the model~(\ref{mf_mod}) in the vectorial
notation as
\begin{align}
\mu(\btheta,\bmath f,t) &= \btheta \cdot \bvarphi(\bmath f,t), \nonumber\\
\bvarphi(\bmath f,t) &= \{\cos(2\pi f_i t), \sin(2\pi f_i t) \}_{i=1,2,\ldots,n}, \nonumber\\
\btheta &= \{A_i,B_i\}_{i=1,2,\ldots,n}, \quad \bmath f = \{f_1,f_2,\ldots,f_n\}.
\end{align}
Now we can eliminate the linear parameters $\btheta$ by means of the linear least-square
regression (that is, to estimate them on the basis of the input data $x_i$, $t_i$, and
$\sigma_i$). To do this, we must solve the following minimization task:
\begin{equation}
\chi^2(\btheta,\bmath f) = \langle (x-\mu)^2 \rangle \longmapsto \min_\btheta
\label{leastsq}
\end{equation}
Here we have borrowed from \citep{Baluev08a} the notation $\langle F(t) \rangle$, which
stands for the weighted sum of the values $F(t_i)$ with the weights $w_i=1/\sigma_i^2$.

Since this $\chi^2$ function is quadratic in $\btheta$, the necessary minimization can be
done elementary:
\begin{align}
\min_\btheta \chi^2(\btheta,\bmath f) &= \langle x^2 \rangle - \langle x \bvarphi \rangle^\mathrm{T} \langle \bvarphi\otimes \bvarphi \rangle^{-1} \langle x \bvarphi \rangle,\nonumber\\
\btheta^* = \arg\min_\btheta \chi^2(\btheta,\bmath f) &= \langle \bvarphi\otimes \bvarphi \rangle^{-1} \langle x \bvarphi \rangle,
\label{chisqmin}
\end{align}
where `$\otimes$' is the dyadic product of vectors ($\bmath a \otimes \bmath b$ is a
matrix constructed of the elements $a_i b_j$). By analogy with the Lomb-Scargle
periodogram, the multi-frequency periodogram, associated to the model~(\ref{mf_mod}), can
be defined as the half of the maximum decrement in $\chi^2$ implied by~(\ref{chisqmin}):
\begin{equation}
z(\bmath f) = \frac{1}{2} \left(\langle x^2 \rangle - \min_\btheta \chi^2 \right) = \frac{1}{2} \langle x \bvarphi \rangle^\mathrm{T} \langle \bvarphi\otimes \bvarphi \rangle^{-1} \langle x \bvarphi \rangle
\label{mf_def}
\end{equation}
Since this $z(\bmath f)$ is basically a test statistic, its large values indicate that the
data probably contain a variation that can be expressed (or at least approximated) by the
model~(\ref{mf_mod}). However, it must be remembered that large values of $z(\bmath f)$
does not yet mean that \emph{all} sinusoidal components of~(\ref{mf_mod}) are actually
present. When the data contain only a single sinusoid at a given frequency $f^*$, the
function $z(\bmath f)$ will be large when $f_i\approx f^*$ only for a single index $i$
(for all $f_j$ with $j\neq i$). In fact, the multi-frequency periodogram can also reveal
single periodicities too, although its detection power in this case would be smaller than
in the single-frequency framework (see Sect.~\ref{sec_sens}). When the data contain $n$
periodicities at $f_i^*$, the multi-frequency periodogram has a multidimensional
cross-like shape: large values at the orthogonal lines $f_i\approx f_i^*$ accompanied by
an especially high peak at the intersection point. In practice, this picture may be of
course made more complicated, e.g. due to the aliasing.

There are pecularities of~(\ref{mf_def}) at the diagonals $f_i=f_j$, where the
model~(\ref{mf_mod}) becomes formally degenerate. This degeneracy can be easily
eliminated, however. For example, for the double-frequency case this can be done by means
of an equivalent replace of the base $\bvarphi$ by the following set:
\begin{align}
\cos(\pi(f_1+f_2) t) \cos(\pi (f_2-f_1) t), \nonumber\\
\sin(\pi(f_1+f_2) t) \cos(\pi (f_2-f_1) t), \nonumber\\
\cos(\pi(f_1+f_2) t) \sin(\pi (f_2-f_1) t)/(f_2-f_1), \nonumber\\
\sin(\pi(f_1+f_2) t) \sin(\pi (f_2-f_1) t)/(f_2-f_1).
\end{align}
It is not hard to check that an arbitrary linear combination of these new base functions
can be transformed to the form of the original model~(\ref{mf_mod}) with $n=2$, but now at
the line $f_1=f_2$ we have a non-degerate base $\{\cos 2\pi f t, \sin 2\pi f t, t \cos
2\pi f t, t \sin 2\pi f t\}$. This also means that a large diagonal value of $z(\bmath f)$
indicates, in general, a \emph{modulated} periodicity with a slowly varying amplitude and
phase.

It is possible to further generalize the definition~(\ref{mf_def}) to deal with some
underlying variation in the data that is deemed to always exist, even if the
multi-periodic signal that we seek is absent. For example, in practice we must always take
into account at least an arbitrary constant offset of $x_i$. We will discuss the ways to
further generalize the multi-frequency periodogram in Section~\ref{sec_gen}. Though some
of these generalizations should be treated in practice as mandatory, we do not write down
the extended definitions here. This is because the definitions here are mainly intended
for the use in Section~\ref{sec_conf} below, where we need to deal with more simple
formulae.

The matrix $\langle \bvarphi\otimes \bvarphi \rangle$ in~(\ref{mf_def}) is the Fisher
information matrix associated to the parameters $\btheta$. In the case of the Lomb-Scargle
periodogram, the single non-diagonal element of this matrix could be made zero by means of
choosing a suitable time shift. This simplified the final formula to a sum of two squared
terms. In the case of double-frequency periodogram this diagonalization is much harder. As
explained e.g. in \citep{SchwCzerny98b}, in the general case we may perform a Gram-Shmidt
orthogonalization of the base $\bvarphi$ in the sense of the scalar product $(a(t),b(t)) =
\langle a(t) b(t) \rangle$. After this procedure, the Fisher matrix will appear strictly
diagonal, so that the periodogram~(\ref{mf_def}) will be expressed as a sum of $2n$
squares. This orthogonalization must be done anew for each new set of frequency values.

The main difference of the multi-frequency periodogram~(\ref{mf_def}) from the
Lomb-Scargle one is in the number of its arguments: it depends on many frequencies, rather
than on only one. Therefore, its visual representation is a multi-dimensional field rather
than a usual graph of a function of a single argument. The honest computation of $z(\bmath
f)$ already for $n\geq 3$ on a full multi-dimensional grid of $f_i$ is a challenge.
However, in practice it might be enough to compute the multi-frequency periodogram only in
the vicinities of a small number of candidate frequencies that are revealed as peaks on
the \emph{single}-frequency periodogram. As we have discussed above, the multi-frequency
periodogram has no isolated peaks; its peaks are located in the intersection nodes of the
grid generated by the mentioned system of candidate frequencies.

It is possible that more computationally fast FFT-like methods of evaluation of $z(\bmath
f)$ may be constructed, similar to the ones already developed for the Lomb-Scargle
periodogram and its other extensions \citep[e.g.][]{Palmer09}. We leave this question
without attention here, since we further focus only on the statistical characteristics of
the multi-frequency periodogram.

\section{Statistical issues coming from the signal multiplicity}
\label{sec_mot}
The construction of the multi-frequency periodogram is not the main goal of our present
paper. This task on itself is rather easy, and a statistic similar to~(\ref{mf_def}) have
already been introduced in the literature; e.g. it was suggested by \citet{Foster95} in
his CLEANest algorithm. Remarkably, the ``global'' version of the CLEANest just utilizes
the direct evaluation of our multi-frequency periodogram (possibly, in some restricted
domains of the entire frequency space).

The goal that we are trying to reach in our work here is the more rigorous treatment of
the statistical significance of the periodicities that we extract from the data.

In practice, a sequential approach is usually adopted to detect the periods in the data:
plot a single-frequency periodogram, find a candidate period, ensure that it is
significant, remove the relevant variation from the data. This sequential approach has two
weaknesses. The first issue appears because for $n$ signal components we must carry out a
complete multiple hypothesis testing procedure rather than to just test each of these $n$
components individually. After we have obtained many of the periods, we have done many
statistical decisions. This means that we have a proportionally larger chance to make a
false detection, in comparison with the extraction of only a single variation. Therefore,
even if each of the period extracted had its $\FAP$ at some tolerable level, say $0.01$,
the overall $\FAP$ for the whole set of the variations is definitely greater: e.g. $\sim
0.1$, if we have claimed to detect ten periods. In the end, although each of these peaks
have passed the settled $\FAP$ threshold individually, we are still unsure about the
reality of all of them.

The other issue of the sequential approach is that we make a rather implicit assumption
that all periods that we have extracted before the given step do actually exist. The
$\FAP$ of the next detected period does not involve the uncertainty related to the very
existence of these previously detected peaks. In practice, we often dealt with the case
when the LS periodogram of the data contain two similar peaks that have only a moderate
significance. Such peaks are not nececcarily aliases of each other, so each of them might
represent a true periodicity. However, in the single-frequency framework, we cannot
rigourously evaluate their significance, because we cannot be entirely sure that one of
these periods is true. If these peaks are similar to each other and pass the $\FAP$
threshold both, we may only conclude that at least one of the relevant periodicities
exists, but to confirm the existence of the both, we have to make an unjustified
assumption that either first or the second one is true. In the end, this leads to an
incorrect $\FAP$ estimation.

To solve the two issues described above we need a method of calculation of the cumulative
significance for a \emph{group} of the periods, in addition to the significances of the
individual periods of this group. In general, the significance of a group may be greater
as well as smaller than the individual significances, because there are two counter-acting
effects. First, the increased number of free model parameters, describing the group of
periodicities, leads to larger noise levels, and this decreases the group significance.
Secondly, the contributions from really existing variations are accumulated when they are
treated jointly, and this increases the group significance.

For example, considering the case of two equal periodogram peaks, four distinct outcomes
are possible:
\begin{enumerate}
\item None of the peaks is significant individually, and they are insignificant as a
couple too.
\item The peaks are significant individually, but not as a couple.
\item The peaks are insignificant individually, but they are significant as a couple.
\item The peaks are significant individually, as well as a couple.
\end{enumerate}
The conclusions following from the cases~I and~IV are obvious: our peaks are just
insignificant or just significant both. In the case~II we would conclude that only one of
the peaks probably exists, but there is no enough observational evidence to confirm that
both of them are true. If the peaks are equal, we cannot decide which of them is the true
one. In the case~III we would draw basically the same conclusion: at least one or even
both of the periods probably exist. We still cannot claim for sure that \emph{both} peaks
are true, since their individual significances are not enough for that.

\begin{figure}
\includegraphics[width=84mm]{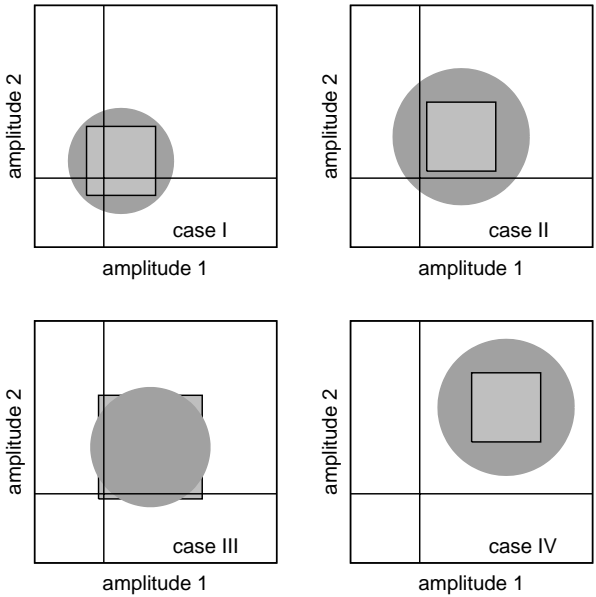}
\caption{The illustration of various types of the interaction between the
single-dimensional individual uncertainties (shown as a square error box) and the
two-dimensional uncertainty domain (shown as a error circle). See text for the detailed
discussion.}
\label{fig_dec}
\end{figure}

We give a graphical illustration of these types of outcomes in Fig.~\ref{fig_dec}. In
these plots we schematically show the uncertainty domains for the components amplitudes
$K_1$ and $K_2$. For the sake of the simplicity, we adopt that the distributions of $K_i$
are uncorrelated and Gaussian, although this is far from the truth. In this illustration,
to ``detect'' a single component or the couple means to ensure that a value $K_i=0$ for a
given $K_i$ is outside of the relevant uncertainty domain. Notice that the uncertainty
segments for a single $K_i$ are not just projections of the uncertainty circle for the
couple $(K_1,K_2)$, because of the different number of the degrees of freedom. The sizes
of the two-dimensional error ellipse and the error box inferred by the single-dimensional
uncertainties is variable and depends on the distributions shape, on the dimnensionality
of the problem, and on the $\FAP$ levels involved. The geometric relastionship between
these regions may be different, although the diameter of the circle is always greater than
the side of the box.

Of course, the cases~II and~III are mostly paradoxical. In the case~II, the both
variations are significant on themselves, but their joint significance is insufficient due
to the dimensionality penalty. Here the circle encompasses the box entirely. In the
case~III, we can detect the components jointly, thanks to the accumulation of their
contributions, but when we treat them individually, we deal with only a portion of this
joint contribution, and this portion appears insignificant. Here the box and consequently
the circle partly cover the axes $K_i=0$, but the origin $K_1=K_2=0$ is still outside of
the circle.

One may think that these subtle geometric and probability effects are rather insignificant
and might be just neglected in practice. This might seem so in the two-dimensional graphs
shown above, but when the number of the frequency component grows, the role of these
effects can become only more importaint. The dimensionality penalty increases, and the
behaviour near the boundary of the uncertainty domain becomes decisive (recall that the
most of the volume of a highly-dimensional ball resides near its surface, rather than in
its core).

By this point, the reader might feel a bit confused, since the issue that we describe
still may look rather fuzzy and foggy. Now he is prepared to look at it from yet another
point of view. \citet{Foster96a,Foster96b} clearly explains how it is important to
explicitly settle the null hypothesis of the signal detection task. This is the hypothesis
that explains the data through a more simple model that does not contain the putative
signal or some its portions. The main difficulty with this null hypothesis is that it
frequently resides in the subconscious domain of a researcher's mind, and is not
explicitly realized even when it is pretty complicated. So, what null hypothesis we
usually bear in mind when we claim ``our dataset contains $n$ periodic components at the
frequencies $f_1,f_2,\ldots,f_n$''?

A traditional detection algorithm, based on the sequential extraction of these components,
assumes that the null hypothesis represents one of the following branches:
\begin{enumerate}
\item No signal at all, only a constant
\item The signal contains a single component $f_1$
\item Signal = component $f_1$ + component $f_2$ and no more
\item Signal = component $f_1$ + component $f_2$ + component $f_3$ and no more

$\ldots$

\item Signal = sum of $n-1$ components, $f_1$ to $f_{n-1}$ and no more
\end{enumerate}
However, this null hypothesis is obviously inappropriate, because it relies on a
particular detection sequence: first extract $f_1$, secondly $f_2$, and so on. In such a
case, our alternative hypothesis, which is a completion of the null one to the entire
space of all admissible models, would embed the following special occurrences: ``signal =
a single component $f_2$'', ``signal = component $f_2$ + component $f_3$ and no more'',
and so on. These particular models are not the ones that we would like to have inside our
alternative hypothesis, since they admit that some of the components that we claimed to
reveal may not exist in some layout. Thus, even if we prove that this alternative
hypothesis is true, this is not the initial proposition we aimed to prove. This occurs
because our original null hypothesis listed above is incomplete. Note that the frequency
values $f_i$ are not allowed to vary arbitrarily here, because they should always reside
inside the relevant periodogram peaks, according to their detection sequence.

Since we want to prove that \emph{each} of the components exists, we must adopt the
\emph{complete} null hypothesis, containing the following model branches:
\begin{enumerate}
\item No signal at all, only a constant
\item The signal contains a single component, any one of $f_i$
\item The signal is a sum of any two (and only two) components $f_i$ and $f_j$
\item The signal is a sum of any three (and only three) components $f_i$, $f_j$, $f_k$

$\ldots$

\item Signal = sum of $n-1$ of the components that exclude any of $f_i$
\end{enumerate}
This new null hypothesis takes into account the possiblity that e.g. the \emph{first}
extracted periodicity is due to the noise, while all others are still true. The previous
null hypothesis undeservedly neglected such possibilities, disregarding their importance
in the case when all relevant periodogram peaks have only a moderate significance, and no
peak clearly dominates.

Our alternative hypothesis, that we want to prove, remains always the same: it states
``the signal is indeed a sum of all $n$ detected components''. Different branches of our
complete null hypothesis thus generate different test statistics. Following the same
enumeration sequence as in the above list, they are:
\begin{enumerate}
\item The maximum of a single available $n$-frequency periodogram
\item $n$ maxima of the $(n-1)$-frequency periodograms (with one of $f_i$ moved to the
base model)
\item $n(n-1)/2$ maxima of $(n-2)$-frequency periodograms (moving some pair of frequencies
to the base model)

$\ldots$

\item $n$ maxima of single-frequency periodograms, for which we have only one of $f_i$ in
the signal model with the rest in the base one
\end{enumerate}
From the computational point of view this task is much easier than it may seem. In
practice, it is enough to maximize all these periodograms in only small vicinities of the
estimated frequency values $f_i$. There is a little chance that the multi-frequency
periodogram will show a remarkable peak at different frequencies. When the data sampling
produce aliases, we may need to also scan the vicinities of all possible alias frequencies
too (since we do not know in advance, which of the peaks are aliases). But in any way, we
do not have to scan the entire multi-frequency space.

The number of the tests involving a $k$-component signal is $C_n^k$ (the binomial
coefficient), hence their total number equals to $\sum_{k=0}^{n-1} C_n^k = 2^n-1$. Each
test generates its own value of the $\FAP$.\footnote{Even though a multi-frequency
periodogram may be maximized in a limited frequency domain around $f_i$ to save
computational resources, its $\FAP$ estimation must always assume the widest domain~---
the Cartesian power of the original frequency scan range. This is because we did not knew
the values of $f_i$ in advance.} In the frequentist framework that we adopt here, there is
always some well-defined \emph{true} signal model, determining which branch of the null
hypothesis is real (when calculating the $\FAP$ we must not admit a thought that the null
hypothesis itself might be wrong). Therefore, only one of the mentioned $2^n-1$ tests
should be relevant. However, since we do not know which one that should be, we can select
a
\emph{worst-case} result, i.e the \emph{maximum} among these $2^n-1$ $\FAP$ values.

In pracice it is a frequent case when the periodograms contain more suspicious peaks than
the ones that our sequential detection algorithm have extracted. Then it might be
necessary to check whether these additional peak provide a better fit of the data in some
other multiple combination. This is essentially what \citet{Foster95} calls the global
CLEANest algorithm. This activity may result in a few of alternative sets of $f_i$.
However, to rigorously prove the statistical significance of each such frequency set, i,e.
to confirm a high confidence probability associated to the proposition ``\emph{each} of
$f_i$ does exist'' (conditionally to the selected rival configuration), we must ensure the
individual significance of $f_i$ (as inferred by the single-frequency periodograms), as
well as their significance in various multi-frequency combinations (within the adopted
configuration of $f_i$). If just one of the relevant $\FAP$ values is too large, we have
to admit that some of the extracted components still might be false positives.

We do not consider here the issue of distinguishing between different alternative sets of
$f_i$, when these sets do not encompass each other (this may appear due to the aliases).
This issue should be treated by means of different statistical tests that are designed to
deal with non-nested models, see \citep{Baluev12}.

There is also a minor matter that we need to highlight. Note that the single-frequency
periodograms that provide some of the test statistics for verifying $f_i$, are the ones
for testing $n$ signals against $n-1$ ones. These ``verification periodograms'' are
different from the single-frequency periodograms that appear during the detection
sequency, in which the hypothesis of $m$ signals is tested against the one with $m+1$
components, subsequently for $m=0,1,\ldots,n-1$ (this is what \citet{Foster95} calls as
the SLICK spectrum). Only the last detection periodogram simultaneously belongs to the
family of the verification ones. Obviously, the verification periodograms should be more
reliable, because they likely involve a more adequate data model. They cannot be used at
the detection stage, since we do not know in advance the number of the components to
extract. We expect that these verification periodograms should typically generate higher
peaks and concequently infer larger individual significances of $f_i$ than the detection
periodograms (though we do not expect a big difference in the $\FAP$ as an abstract
function of the maximum peak height $z$). This effect is nonetheless counterbalanced by
the need to ensure that all multi-frequency combinations are also significant.

Therefore, to properly detect multiple periodic signals, the most of the significance
tests that we apply, should be multi-frequency ones. Construction of a method that would
allow to estimate such a joint multi-frequency significance is the main goal of our work.

\section{Asymptotic significance levels}
\label{sec_conf}
As it is well-known, the false alarm probability tied to an observed value of a signal
detection statistic, is related to the distribution function of this quantity. When the
frequencies $f_i$ are fixed (known a priori), such decisioning quantity would be just the
value $z(\bmath f)$. In this case, the model~(\ref{mf_mod}) would be entirely linear, and
hence $2z(\bmath f)$ would follow the chi-square distribution with $2n$ degrees of
freedom. For $n=2$ this would imply $\FAP(z) = e^{-z} (z+1)$, for instance. However, in
practice the frequencies $f_i$ are unknowns like $\btheta$. Since $f_i$ are non-linear,
the maximization of the periodogram $z(\bmath f)$ is a non-trivial task, as well as the
calculation of the necessary distribution function of its \emph{maximum} peaks. To solve
this task we will use the approach described by \citet{Baluev13b}. This approach is based
on the generalized Rice method for random fields presented by \citet{AzaisDelmas02}. The
Rice method allows to obtan an estimation of the necessary false alarm probability in the
following form:
\begin{equation}
\FAP(z) \lesssim M(z),
\label{rice-est}
\end{equation}
where $z$ is the observed periodogram maximum, and $M(z)$ is an explicitly-defined
function.

The high practical value of the estimation~(\ref{rice-est}) is founded on the following
things: (i) it usually has a good or at least satisfactory accuracy (in terms of the
difference $\FAP-M$ or at least in terms of the $z$-level thresholds that are mapped to a
given value of $\FAP(z)$ or $M(z)$); (ii) this accuracy increases for larger $z$, which
has more practical importance, since in practice we need to have a good accuracy mainly
for the small $\FAP$s; (iii) in the case when the deviation between $M(z)$ and $\FAP(z)$
is too large, the function $M(z)$ still serves as a majorant for $\FAP(z)$, guaranteeing
that the number of false detection will never exceed the desired small level; (iv) the
function $M(z)$ often can be approximated by a simple and accurate elementary formula.

For the Lomb-Scargle periodogram, for example, we obtained in \citep{Baluev08a}:
\begin{equation}
\FAP(z) \lesssim M(z) \approx W e^{-z} \sqrt z.
\label{fap_sfreq}
\end{equation}

Now, let us first consider the more easy case of the double-frequency periodogram, and
after that we proceed to dealing with the general multi-frequency periodogram.

Although the method in \citep{Baluev13b} was originally designed to deal with a single
(though non-sinusoidal) periodicity to detect, it still can be applied to a
multi-frequency case with a help of a ruse. We cannot just directly substitute the
model~(\ref{mf_mod}) in the formulae from \citep{Baluev13b}, because~(\ref{mf_mod})
contains $2n$ linear coefficients, instead of only a single signal amplitude, as we need
for \citep{Baluev13b}. We need to transform~(\ref{mf_mod}) to an equivalent form, possibly
looking more complicated, but containing only a single common amplitude parameter. For
$n=2$, one way to do so leads to the following representation:
\begin{equation}
\mu = K h, \quad h = \cos\alpha \cos(2\pi f_1 t + \lambda_1) + \sin\alpha \cos(2\pi f_2 t + \lambda_2),
\label{df_mod2}
\end{equation}
where $K$ is the mentioned single amplitude, and $h$ is a function of the time $t$ and of
the five free parameters, including the new auxiliary non-linear parameter $\alpha$,
responsible for the mixture of the two sinusoidal components.

When deriving our further results we will use the approximation of the `uniform phase
coverage' (UPC), as we called it in \citep{Baluev13b}. In this approximation, we neglect the
quantities like
\begin{equation}
\langle t^k \cos\omega t \rangle, \quad \langle t^k \sin\omega t \rangle,
\label{tksc}
\end{equation}
in comparison with $\langle |t|^k \rangle$ and similar quantities (in our case $k=0$, $1$,
and $2$). This approximation is formally good only when $\omega$ is outside of a peak of
the spectral window (the spectral leakage effect is negligible for a given $\omega$).
However, as we have already discussed and demonstrated in many works
\citep{Baluev08a,Baluev09a,Baluev13b} the presence of the spectral leakage itself does not
yet significantly corrupt the quality of the final UPC approximation of the $\FAP$. This
is because we should eventually perform an integration over a wide frequency range, and
the anomalies generated by narrow peaks of the spectral window appear negligible after
such an integration.

The UPC approximation basically enabled us just to drop all the terms of the
type~(\ref{tksc}) anywhere we met them, leaving only the dominating terms $\langle t^k
\rangle$.

According to \citet{Baluev13b}, first we need to construct from the model $h$ a normalized
function $\psi$, such that $\langle\psi\rangle\equiv 0$, and $\langle\psi^2\rangle\equiv
1$. Since under the assumption of UPC we have $\langle h^2\rangle\approx 0$ already and
$\langle h^2\rangle\approx \langle 1\rangle/2$, we can put $\psi \approx h \sqrt{2/\langle
1\rangle}$. Finally we need to evaluate the matrix $\mathbfss G = \langle \psi' \otimes
\psi' \rangle$, based on the gradient $\psi' \approx h' \sqrt{2/\langle 1\rangle}$. The
gradient of $h$ looks like:
\begin{align}
h'_\alpha &= -\sin\alpha \cos(2\pi f_1 t + \lambda_1) + \cos\alpha \cos(2\pi f_2 t + \lambda_2), \nonumber\\
h'_{\lambda_1} &= -\cos\alpha \sin(2\pi f_1 t + \lambda_1), \nonumber\\
h'_{f_1} &= -2\pi t \cos\alpha \sin(2\pi f_1 t + \lambda_1), \nonumber\\
h'_{\lambda_2} &= -\sin\alpha \sin(2\pi f_2 t + \lambda_2), \nonumber\\
h'_{f_2} &= -2\pi t \sin\alpha \sin(2\pi f_2 t + \lambda_2).
\end{align}
Using these formulae and UPC approximation, we obtain
\begin{align}
\mathbfss G &\approx
\left(\begin{array}{@{}c@{\;}c@{\;}c@{\;}c@{\;}c@{}}
 1 & 0 & 0 & 0 & 0 \\
 0 & \cos^2\alpha & 2\pi \bar t \cos^2\alpha & 0 & 0\\
 0 & 2\pi \bar t \cos^2\alpha & 4\pi^2 \bar{t^2} \cos^2\alpha & 0 & 0 \\
 0 & 0 & 0 & \sin^2\alpha & 2\pi \bar t \sin^2\alpha \\
 0 & 0 & 0 & 2\pi \bar t \sin^2\alpha & 4\pi^2 \bar{t^2} \sin^2\alpha \\
\end{array}\right), \nonumber\\
&\overline{t^k} = \langle t^k \rangle/\langle 1\rangle,
\end{align}
which also implies
\begin{equation}
\sqrt{\det\mathbfss G} \approx \frac{\pi}{4} T_\mathrm{eff}^2 \sin^2 2\alpha, \qquad T_\mathrm{eff} = \sqrt{4\pi (\overline{t^2}- {\overline t}^2)}.
\label{matrG}
\end{equation}
We note that the quantity $T_\mathrm{eff}$, emerging here, is the effective length of the
time series that was first introduced in \citep{Baluev08a}.

The quantity $\sqrt{\det\mathbfss G}$ should now be integrated over the space of all five
free parameters $\alpha, \lambda_i, f_i$ to obtain the final result:
\begin{align}
\FAP(z) &\lesssim M(z) \simeq A e^{-z} z^{d/2-1}, \nonumber\\
A &= \frac{1}{2\pi^{d/2}} \int \sqrt{\det\mathbfss G}\; df_1 df_2 d\lambda_1 d\lambda_2 d\alpha,
\label{fap}
\end{align}
where $d$~--- the number of free model parameters~--- is now equal to $6$.

Here we should take care of one subtle thing: over what exactly domain we must do the
integration? We must take into account that the mentioned parameters of the signal satisfy
a few relations of equivalence. Namely, the following six vectors of the parameters
\begin{equation}
\begin{array}{@{\{}rrrrrr@{\},}}
K, & \alpha, & \lambda_1, & f_1, & \lambda_2, & f_2 \\
-K, & \alpha, & \lambda_1+\pi, & f_1, & \lambda_2+\pi, & f_2 \\
-K, & \alpha+\pi, & \lambda_1, & f_1, & \lambda_2, & f_2 \\
K, & \pi-\alpha, & \lambda_1+\pi, & f_1, & \lambda_2, & f_2 \\
K, & -\alpha, & \lambda_1, & f_1, & \lambda_2+\pi, & f_2 \\
K, & \frac{\pi}{2}-\alpha, & \lambda_2, & f_2, & \lambda_1, & f_1\\
\end{array}
\label{parequiv}
\end{equation}
all describe the same signal~(\ref{df_mod2}). To encompass all possible signals inside a
frequency range $[f_\mathrm{min},f_\mathrm{max}]$, simultaneously throwing away all the
duplicates, we may consider the following domain:
\begin{equation}
K \geq 0,\quad 0 \leq\alpha\leq \frac{\pi}{4}, \quad 0 \leq \lambda_i \leq 2\pi, \quad f_\mathrm{min} \leq f_i \leq f_\mathrm{max}.
\label{dfp_domain}
\end{equation}
Notice that the condition $K\geq 0$ is exactly the one required in \citep{Baluev13b}
for~(\ref{fap}) to be valid (otherwise $A$ should be doubled). The condition $\alpha\leq
\pi/4$ appears because due to the last equivalence of~(\ref{parequiv}) the replacement
$\alpha \mapsto \pi/2-\alpha$ would just swap the signal components with each other,
keeping the sum intact. Performing the integration over the domain described, we
eventually obtain
\begin{equation}
\FAP(z) \lesssim M(z) \approx \frac{\pi}{16} W^2 e^{-z} z^2, \quad W = T_\mathrm{eff} (f_\mathrm{max}-f_\mathrm{min}).
\label{fap_dfreq}
\end{equation}

The formula~(\ref{fap_dfreq}) is valid for the case when the frequencies $f_i$ belong to
the same range. Sometimes we may have some prior information that would imply different
ranges for $f_1$ and $f_2$. In the case when these ranges do not intersect, we should
extend the integration domain from $0\leq \alpha\leq \pi/4$ to $0\leq \alpha\leq\pi/2$,
because now the frequency components are not freely swappable and the last equivalence
of~(\ref{parequiv}) is no longer valid. This will double the result. In the most general
case, when the frequency ranges are partially intersecting, we may write down:
\begin{equation}
\FAP(z) \lesssim M(z) \approx \frac{\pi}{16} (2 W_1 W_2 - W_{12}^2) e^{-z} z^2,
\label{fap_dfreq_gen}
\end{equation}
where $W_1$ and $W_2$ are associated with the frequency ranges of $f_1$ and $f_2$, while
$W_{12}$ is related to their common intersection.

The formulae for $M(z)$ in~(\ref{fap}) and, consequently~(\ref{fap_dfreq})
and~(\ref{fap_dfreq_gen}), make some additional approximating assumptions that we still
need to discuss. The first thing it neglects is the effect of the
domain~(\ref{dfp_domain}) boundary, which importance was described in \citep{Baluev13b}.
In the case of the double-frequency periodogram, the boundary sides
$f_1=f_{\mathrm{min},\mathrm{max}}$ and $f_2=f_{\mathrm{min},\mathrm{max}}$ generate an
extra term in~(\ref{fap_dfreq}) of the order of $\sim W e^{-z} z^{3/2}$, and an extra term
for the vertices of the relevant frequency box would be $\sim e^{-z} z$. The non-frequency
parameters $\lambda_i$ and $\alpha$, thanks to their periodicity, do not generate any
boundary effects. Anyway, all these extra terms are negligible, because the value of $W$
in practice is typically large or very large ($\sim 100$ or $\sim 1000$ or even more). The
relevant correction to~(\ref{fap_dfreq}) would have a very small relative magnitude of
$\sim 1/(W\sqrt z)$ and $\sim 1/(z W^2)$. No doubts, it can be safely neglected in
practice.

The other small terms that were dropped off in~(\ref{fap}), have the relative magnitude of
$\sim 1/z$ and $\sim 1/z^2$. As we have already discussed in \citep{Baluev13b}, these
terms are usually very difficult to evaluate, because they involve manipulations already
with second-order dervatives of $\psi$, combined in tensors of order $4$ and dimension
$n$. In our case, it is a $5\times 5\times 5 \times 5$ tensor, for instance. These terms
are also expected to be negligible, because we are usually interested in large values of
$z$: typically, when $z$ is smaller than $10$, the signal is very uncertain, and the
associated false alarm probability is large, so we just have no real need to know this
probability with a good precision. However, in the particular case of the double-frequency
periodogram, we were able to rigorously evaluate these terms, rather than just to blindly
neglect them. This appeared possible because of the simplicity of the signal model.
According to our results, the corrected general expression~(\ref{fap_dfreq_gen}) looks
like
\begin{align}
\FAP(z) &\lesssim M(z) \approx e^{-z} \Bigg[\frac{\pi}{16} (2W_1 W_2 - W_{12}^2) (z^2 + 2z + 2) - \nonumber\\
 & - W_1 W_2 \left(z + \frac{1}{2}\right) \Bigg].
\label{fap_dfreq_gen2}
\end{align}
In the case of the same frequency range for the both frequencies ($W_1=W_2=W_{12}=W$) we
therefore have
\begin{equation}
\FAP(z) \lesssim M(z) \approx W^2 e^{-z} \left[\frac{\pi}{16} z^2 + \left(\frac{\pi}{8}-1\right) z + \frac{\pi-4}{8} \right].
\label{fap_dfreq2}
\end{equation}

We do not give the detailed derivation of~(\ref{fap_dfreq_gen2}) and~(\ref{fap_dfreq2}),
because it still appeared very complicated. We only mention that we used the general
formulae of Proposition~b of Theorem~1 by \citet{AzaisDelmas02}. Some extra discussion can
be also found in \citep{Baluev13b}. Actually, we wrote down these refined results here
only to demonstrate below that their difference from~(\ref{fap_dfreq_gen})
and~(\ref{fap_dfreq}) is negligible.

Let us now consider the general multi-frequency case. Now we may put
\begin{equation}
\mu = K h, \quad h = \sum_{k=1}^n \nu_k \cos(2\pi f_k t + \lambda_k),
\label{mf_mod2}
\end{equation}
where $\nu_k$ are components of a unit vector $\bnu$ parameterized by $n-1$ spherical
angles forming another vector $\balpha$. Now the gradient of $h$ looks like
\begin{align}
h'_\balpha &= \sum_{k=1}^n \frac{\partial\nu_k}{\partial\balpha} \cos(2\pi f_k t + \lambda_k), \nonumber\\
h'_{\lambda_k} &= -\nu_k \sin(2\pi f_k t + \lambda_k), \nonumber\\
h'_{f_k} &= -2\pi t \nu_k \sin(2\pi f_k t + \lambda_k).
\end{align}
These expressions allow us to write down the matrix $\mathbfss G$ very similarly to the
double-frequency case; the necessary determinant can be then expressed as
\begin{equation}
\sqrt{\det\mathbfss G} \approx \pi^{n/2} T_\mathrm{eff}^n (\nu_1\nu_2\ldots\nu_n)^2 \sqrt{\det\left(\frac{\partial\bnu}{\partial\balpha}^{\rm T}\frac{\partial\bnu}{\partial\balpha}\right)}
\end{equation}
The last multiplier in this expression, containing the gradient of $\bnu$ over $\balpha$,
is rather unpleasant and needs some simplification. Let us define an auxiliary square
matrix $\mathbfss R = (\bnu, \partial\bnu/\partial\balpha)$, and try to find its
squared determinant:
\begin{equation}
(\det\mathbfss R)^2 = \det(\mathbfss R^{\rm T} \mathbfss R) =
\det\left(\begin{array}{@{}cc@{}}
\bnu^{\rm T}\bnu & \bnu^{\rm T} \frac{\partial\bnu}{\partial\balpha} \\
\frac{\partial\bnu}{\partial\balpha}^{\rm T} \bnu & \frac{\partial\bnu}{\partial\balpha}^{\rm T}\frac{\partial\bnu}{\partial\balpha} \\
\end{array}\right).
\label{aux_nu}
\end{equation}
Since the identity $\bnu^2\equiv 1$ holds true for every $\balpha$, the off-diagonal
elements of the last matrix in~(\ref{aux_nu}) are zero, the top-left element is unit, and
we finally obtain
\begin{equation}
\sqrt{\det\mathbfss G} \approx \pi^{n/2} T_\mathrm{eff}^n (\nu_1\nu_2\ldots\nu_n)^2 \left|\det\left(\bnu,\frac{\partial\bnu}{\partial\balpha}\right)\right|.
\label{matrGmult}
\end{equation}
The last determinant basically represents the Jacobian of the transition from the
Cartesian to the spherical coordinate system, i.e. from $\bmath x=r\bnu$ to the pair
$(r,\bnu)$.

Let us recall the formulae of the multi-dimensional spherical parametrization:
\begin{align}
\nu_1 &= \cos\alpha_1, \nonumber\\
\nu_2 &= \sin\alpha_1 \cos\alpha_2, \nonumber\\
\nu_3 &= \sin\alpha_1 \sin\alpha_2 \cos\alpha_3, \nonumber\\
 &\ldots \nonumber\\
\nu_{n-1} &= \sin\alpha_1 \sin\alpha_2 \ldots \sin\alpha_{n-2} \cos\alpha_{n-1}, \nonumber\\
\nu_n &= \sin\alpha_1 \sin\alpha_2 \ldots \sin\alpha_{n-2} \sin\alpha_{n-1},
\end{align}
where all $\alpha_i$ except for the last one, should in general reside in the segment
$[0,\pi]$, and $\alpha_{n-1}$ is allowed to vary inside $[0,2\pi]$.

With these formulae we can rewrite~(\ref{matrGmult}), after rather long but
straightforward manipulations, in a detailed form:
\begin{equation}
\sqrt{\det\mathbfss G} \approx \pi^{n/2} T_\mathrm{eff}^n \prod_{k=1}^{n-1} \sin^{3k-1} \alpha_{n-k} \cos^2\alpha_{n-k}.
\label{matrGn}
\end{equation}

Again the issue arises, what integration domain we should adopt for~(\ref{matrGn})? This
should be the largest possible domain that still does not contain any equivalent pairs of
points (describing the same signal). Now it is easier to consider the case when the
frequencies $f_i$ all belong to different segments, and these segments do not intersect
with each other. Then we may adopt the domain
\begin{equation}
K \geq 0,\quad 0 \leq\alpha_i\leq \frac{\pi}{2}, \quad 0 \leq \lambda_i \leq 2\pi, \quad f_{i,\mathrm{min}} \leq f_i \leq f_{i,\mathrm{max}}.
\end{equation}
The condition $\alpha_i\in [0,\pi/2]$ appeared because the signs of individual terms
in~(\ref{mf_mod2}) are managed by the longitudes $\lambda_i$, while all $\nu_i$ must then
be positive to get rid of duplicate signals in the domain.

Substituting all necessary quantities to~(\ref{fap}) and integrating, we obtain
\begin{align}
\FAP(z) &\lesssim M(z) \simeq A_n e^{-z} z^{3n/2-1}, \nonumber\\
 A_n &= W_1 W_2 \ldots W_n I_1 I_2 \ldots I_{n-1}, \nonumber\\
 I_k &= 2 \int\limits_0^{\pi/2} \sin^{3k-1} \alpha\, \cos^2 \alpha\, d\alpha = B\left(\frac{3k}{2},\frac{3}{2}\right).
\label{fap_mfreq_gen}
\end{align}

For the more practical case, in which there is a single common frequency segment $W$, it
is rather difficult to define the necessary integration domain for $\alpha_i$. Instead,
let us try to correct the formula~(\ref{fap_mfreq_gen}). What is the underlying reason
making the normalization of the $\FAP$ different for the cases of the shared and
independent frequency segments? Of course, this is the symmetry property of the
multi-frequency periodogram. For the double-frequency periodogram the obvious identity
$z(f_1,f_2) = z(f_2,f_1)$ implies that only a half of the frequency square $W\times W$ is
informational; another half is just a mirror copy. This halves the resulting $\FAP$
value~--- compare~(\ref{fap_dfreq}) and~(\ref{fap_dfreq_gen}) with $W_{12}=0$. In the case
of the general multi-frequency periodogram only a single $n$-simplex inside the entire
frequency cube is informational. Since the volume of this simplex constitutes $1/n!$
fraction of the cube volume, we have for this case
\begin{align}
\FAP(z) &\lesssim M(z) \simeq \tilde A_n e^{-z} z^{3n/2-1}, \nonumber\\
 & \tilde A_n = \frac{W^n}{n!} I_1 I_2 \ldots I_{n-1}.
\label{fap_mfreq}
\end{align}

From this general approximation we can easily reconstruct the $\FAP$ formulae for the
Lomb-Scargle periodogram~(\ref{fap_sfreq}) and for the double-frequency
periodogram~(\ref{fap_dfreq}).

We do not consider here the most general case when the frequency segments are different
but may have common parts.

\section{Some further extensions}
\label{sec_gen}
The multi-frequency periodogram defined in~(\ref{mf_def}) assumes the empty null
hypothesis: the input data are expected to represent the pure noise or the pure noise plus
a signal with no even an offset. Therefore, another possible way to generalize the
multi-frequency periodogram is to consider some non-trivial base models describing an
expected underlying variation (the non-trivial null hypothesis). In practice there should
be at least a free constant term in the null hypothesis, because the data $x_i$ almost
always have an arbitrary offset to be determined from the data.

It is already clearly demonstrated in the literature that this is a bad practice to just
pre-center the input time series and then pass the residuals to the LS periodogram
\citep{Cumming99}. Instead, we must honestly perform the linear regression under the null
hypothesis (``data are equal to an unknown constant''), under the alternative one (``the
data are equal to a constant plus the putative signal'') and evaluate the inferred test
statistic. For the $n=1$ (Lomb-Scargle) case this was essentially done by
\citet{FerrazMello81}, who defined the so-called Date-Compensated Discrete Fourier
Transform, DCDFT. This periodogram is known rather well already, though in the literature
it is referred to under different names; e.g. it is called as just ``the generalized
periodogram'' by \citet{ZechKur09}. We prefer an intuitive and concise name ``the
floating-mean periodogram'', given by \citet{Cumming99}.

\citet{Cumming99} also suggested to extend the floating-mean periodogram further, taking
into account an arbitrary linear or a quadratic trend in the data. Moreover, it is quite
easy to construct a generalized periodogram with an arbitrary multi-parametric linear
model (e.g. a polynomial trend) of the underlying variation \citep{Baluev08a}. In this
case, the definition~(\ref{mf_def}) must also involve the linear regression made for this
non-trivial base model. We should replace~(\ref{mf_def}) by
\begin{equation}
z(\bmath f) = \frac{1}{2} \left(\min_{\btheta_{\mathcal H}} \chi^2_{\mathcal H}(\btheta_{\mathcal H}) -
                                \min_{\btheta_{\mathcal K}} \chi^2_{\mathcal K}(\btheta_{\mathcal K},\bmath f) \right),
\end{equation}
where $\chi^2_{\mathcal H,K}$ are the $\chi^2$ goodness-of-fit functions~(\ref{leastsq}),
now corresponding to either null ($\mathcal H$) or the alternative ($\mathcal K$) models.
Previously, for an empty $\mathcal H$, we had just $\chi_{\mathcal H}^2 \equiv \langle
x^2\rangle$.

When the underlying variation is modelled by a low-order polynomial, the theory of
significance levels from Section~\ref{sec_conf} remains practically unchanged, because
under the assumption of UPC the powers of time, $t^k$, appear orthogonal to the sine and
cosine functions, since during the calculation of the extreme-value distribution we anyway
neglect the terms like~(\ref{tksc}).\footnote{The periodogram itself still properly takes
into account the non-orthogonality of $t^k$ to the trigonometric functions; we emphasize
that the UPC approximation is used only to approximate its \emph{distribution}. Get back
to Section~\ref{sec_conf} for a justification.} For the extensions of the Lomb-Scragle
periodogram we checked it in \citep{Baluev08a,Baluev09a}, and below we verify this for the
double-frequency periodogram too, using the Monte Carlo simulations.

In the Sections~\ref{sec_def} and~\ref{sec_conf} we only considered the case when
$\sigma_i$ are known precisely. In practice we usually do not know them with good
accuracy. Usually only the weights $w_i$ of the observations are known, while the full
variances are determined through $w_i$ as $\sigma_i^2=\kappa/w_i$, where $\kappa$ is an
extra unknown parameter. In this case we need to introduce some normalization of the
periodogram, since the value of $\chi^2$ is now proportional to the unknown multiplier
$\kappa$. We recommend to use in this case a multi-frequency analog of the periodogram
$z_3$ from \citep{Baluev08a}. We can define it here as
\begin{equation}
z_3(\bmath f) = -\left(\frac{N_{\mathcal H}}{2}-n\right) \log \left(1-\frac{2 z(\bmath f)}{\min\chi^2_{\mathcal H}}\right),
\end{equation}
where $N_{\mathcal H}=N-\dim\btheta_{\mathcal H}$.

This periodogram is related to the likelihood ratio statistic, and for large $N$ its
distributions (including the distribution of the maximum) are asymptotically the same as
the ones of the $\chi^2$-statistic $z(\bmath f)$. We have already discussed this issue in
\citep{Baluev08b,Baluev13b}. For the single-frequency variant of $z_3$ the reader may find
more accurate expression of the type~(\ref{rice-est}), which do not rely on the
$N\to\infty$ asymptotics, in \citep{Baluev08a}. It is easy to ensure that they are indeed
asymptotically equivalent to the similar expressions for $z$, if the extra condition
$z,z_3 \ll N$ is also satisfied. The principal nature of this condition is explained in
\citep{Baluev08b}, and we believe that it should be valid in the case of the
multi-frequency periodogram too. Unfortunately, at present we cannot generalize the more
accurate formula for $M(z_3)$ from \citep{Baluev08a} to the multi-frequency case, because
this would need the generalized Rice method for non-Gaussian random fields, which is to
our awareness still poorly developed. At present we have to approximate the function
$M(z_3)$ by its analog $M(z)$, as we have just described. Notice that the $\FAP$ will be
anyway extremely small for the values of $z$ or $z_3$ as large as $N$, so in practice the
need to satisfy an extra condition like $z\ll N$ should not produce any significant side
effects. At least, this is well confirmed by the numerical simulations discussed below.

Finally, we may consider a weakly non-linear base model, which may include e.g. some
previously detected periodicities. These periodograms are extensively referred to in
Section~\ref{sec_mot}. The base frequencies are formally non-linear parameters, but they
are linearizable, since once the periodicity is extracted, its frequency always resides
within a narrow periodogram peak. For these periodograms the argumentation of the above
paragraph applies qualitatively: the $\FAP$ formulae of Section~\ref{sec_conf} should work
in the asymptotic sense $N\to\infty$. However, we should realize that in concrete
practical cases with a concrete $N$ these $\FAP$ approximation may fail sometimes, e.g.
after we have already extracted a large number of the signal components. This issue is
something that we leave for future work.

\section{Simulations}
\label{sec_simul}
\begin{figure*}
\begin{tabular}{@{}c@{}c@{}}
\includegraphics[width=0.456\linewidth]{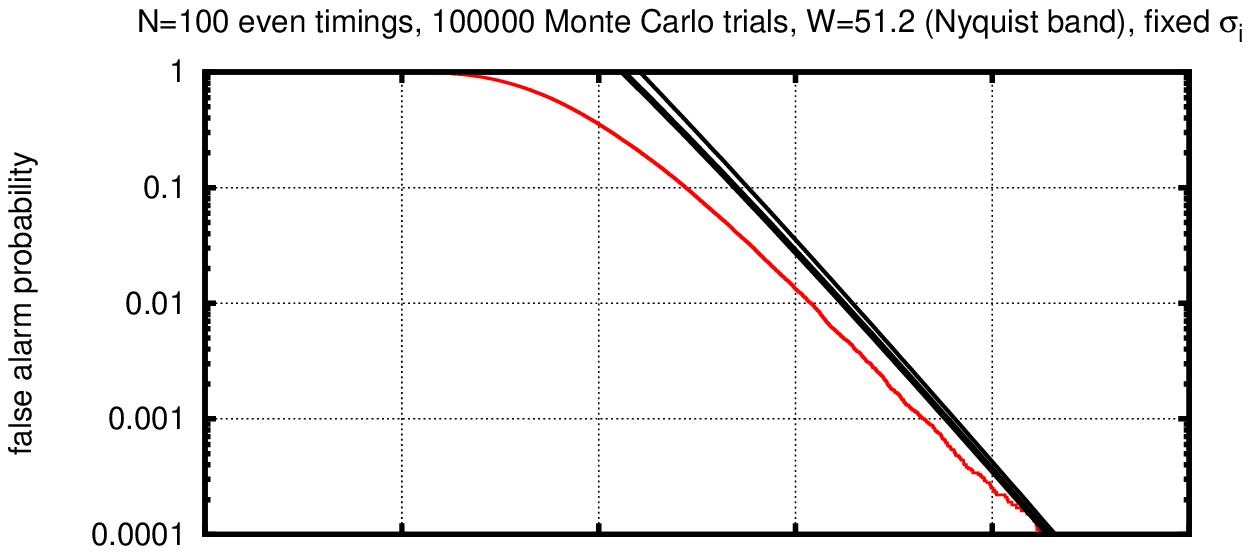} & \includegraphics[width=0.394\linewidth]{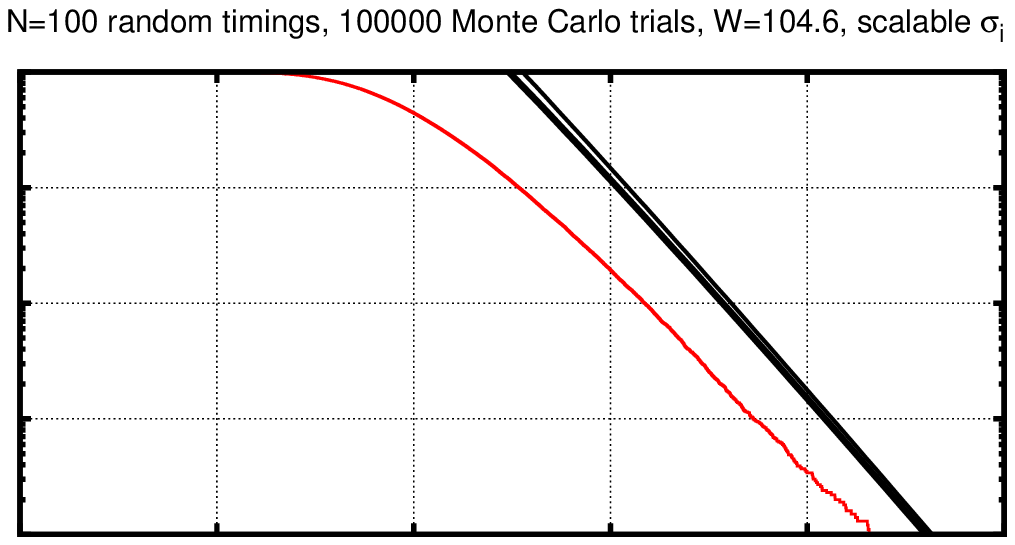}\\
\includegraphics[width=0.456\linewidth]{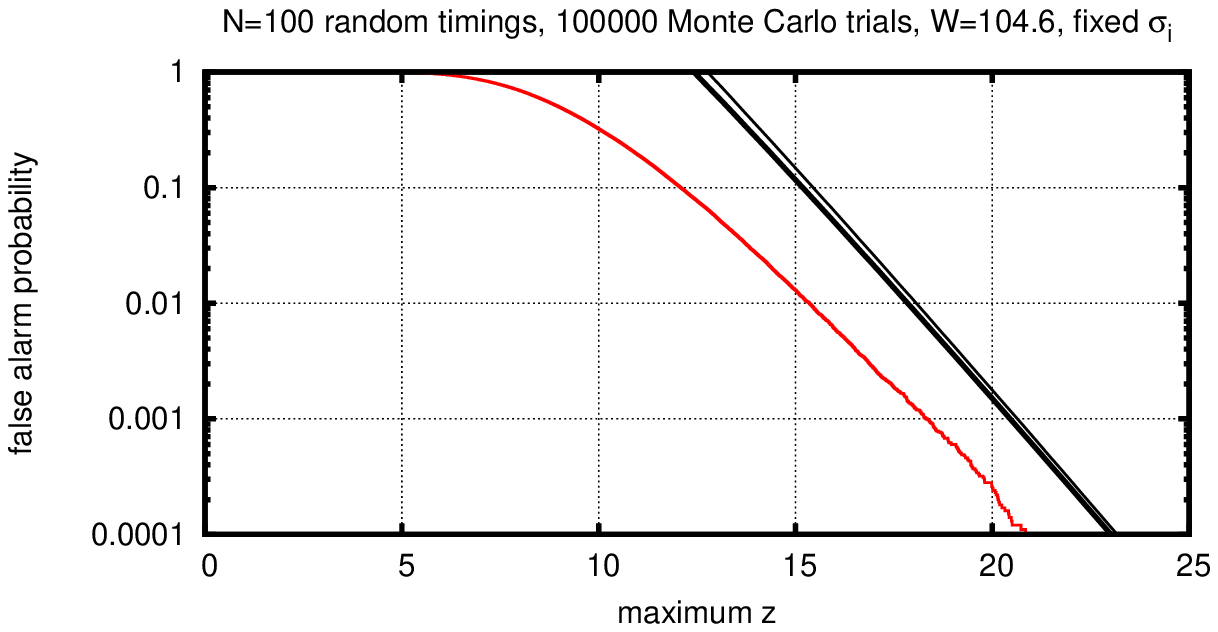} & \includegraphics[width=0.394\linewidth]{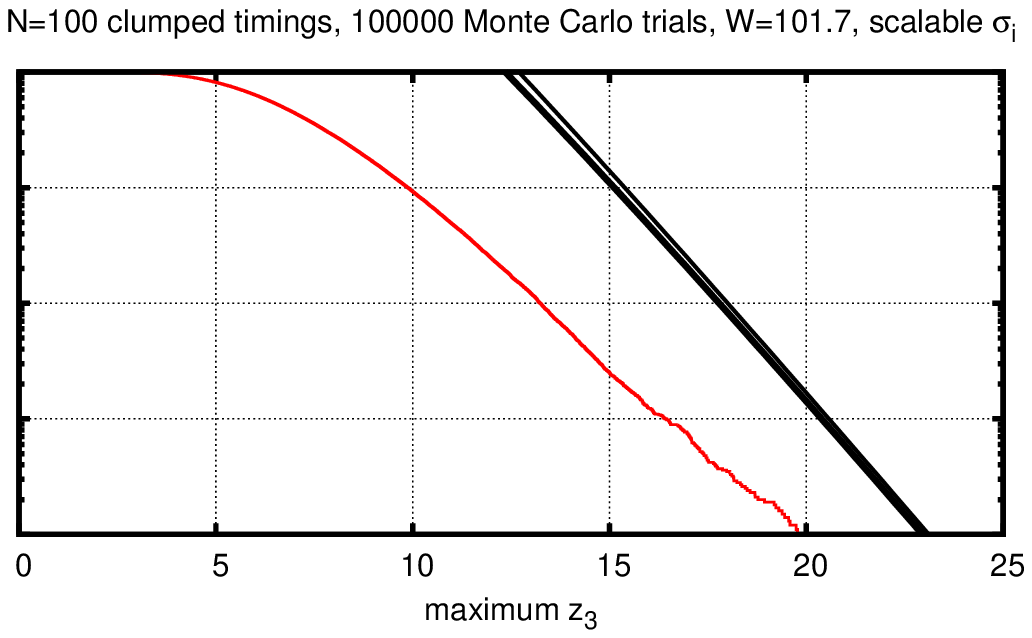}\\
\end{tabular}
\caption{The graphs comparing the analytic $\FAP$ estimations~(\ref{fap_dfreq}) (thick
solid curves) and~(\ref{fap_dfreq2}) (thin solid curves) and the simulated $\FAP$ (noisy
thin curves, red in the electronic version). The panels to the left correspond to the
cases with the noise uncertainties $\sigma_i$ are known a priori; the ones to the right
are for the fixed-weights model with $\sigma_i^2 \propto 1/w_i$. In the case of ``clumped
timings'' (right-bottom panel) the $N=100$ points of the data were equally split in $10$
equidistant groups with $90\%$ time gaps between them. This implies a very strong
aliasing, which invalidates our $\FAP$ formula as an approximation but does not break its
upper limit property.}
\label{fig_simul}
\end{figure*}

We have done some numerical simulations to check our analytical results. In these
simulations, we used the double-frequency periodogram with a free constant term in the
base model (a double-frequency analog of the generalized floating-mean periodogram by
\citet{FerrazMello81} and \citet{ZechKur09}). We considered both the case with known
$\sigma_i$ and the case when $\sigma_i$ contain an unknown factor. In all the cases we
deal with a single range for the both frequencies $f_i$ (meaning that $W_1=W_2=W_{12}=W$).
The main results are shown in Fig.~\ref{fig_simul}, where for $W\approx 100$ and different
time series we plot the simulated $\FAP$ curves together with the analytic
approximation~(\ref{fap_dfreq}) and its refined version~(\ref{fap_dfreq_gen2}). As for the
Lomb-Scargle periodogram, the quality of our analytic formulae is the best for an even
time series, and degrades for a time series with strong aliasing. However, our estimation
always consitutes an upper limits on the $\FAP$. We can always be pretty sure that if we
got the estimation of, e.g. $M(z)=1\%$, the actual value of $\FAP$ may be almost the same
or smaller than this value. This means that the use of $M(z)$ from~(\ref{fap_dfreq})
instead of $\FAP$ does not increase the number of false alarms above the desired level.

A bad side effect of possible deviation between $M$ and $\FAP$ is the increase of the
detection threshold. However, in the worst case of Fig.~\ref{fig_simul} this increase
constitutes the relatitve magnitude of $\sim 1/3$, which is not catastrophic at all.

\begin{figure}
\includegraphics[width=\linewidth]{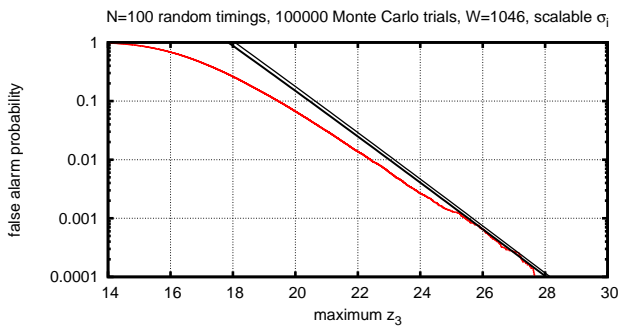}
\caption{Same as the right-top panel of Fig.~\ref{fig_simul}, but for a much wider
frequency range. Notice that the increase of $W$ made our analytic estimations more
accurate.}
\label{fig_simul2}
\end{figure}

It may be noted that with the increase of $W$ to a more practical level of $\sim 1000$,
the precision of our $\FAP$ estimate gets even better (Fig.~\ref{fig_simul2}). We believe
this is because larger value of $W$ makes the signal/noise threshold to move to a higher
$z$-level, and the Rice method becomes more accurate for $z\to\infty$, since it has an
asymptotic nature.

At last, we can conclude, that the refined formula~(\ref{fap_dfreq_gen2}) does not have
any practical advatage with respect to the original more simple
expressions~(\ref{fap_dfreq_gen}) and~(\ref{fap_dfreq}). As we expected, only the leading
$z^2$ term is important in~(\ref{fap_dfreq_gen2}), and the others are neglectable. As
usually, the main source of the error of our approach is the intrinsic error of the Rice
method (the difference between $M(z)$ and $\FAP(z)$), rather than the inaccuracy of the
approximated $M(z)$.

Performing similar Monte Carlo simulations for a multi-frequency periodogram with $n>2$ is
not feasible due to the huge computational demands. Nonetheless, it would not be
comfortable just to leave the multi-frequency $\FAP$ approximation from
Section~\ref{sec_conf} without any numerical verification at all, because the relevant
math manipulations were not trivial. At least, we should ensure that that these $\FAP$
formulae do not hide e.g. a error in the coefficient.

The direct maximization of $z(\bmath f)$ is a computation-heavy procedure, but in some
simple cases it can be dramatically simplified. Namely, when the observations are
distributed uniformly in time (strictly evenly or randomly with uniform distribution), we
can apply the UPC approximation to the periodogram itself, rather than only to its $\FAP$
approximation. In this case the sinusoidal terms in~(\ref{mf_mod}) appear practically
orthogonal to each other. Then, considering that $\sigma_i$ are known, the following
approximate expansion appears:
\begin{equation}
z(\bmath f) \approx z_{\rm sf}(f_1) + z_{\rm sf}(f_2) + \ldots + z_{\rm sf}(f_n).
\label{expand}
\end{equation}
Here the function $z_{\rm sf}$ is the relevant single-frequency periodogram (actually, its
Ferraz-Mello's version in our case). This expansion is valid everywhere except for the
neighbourhoods of the diagonals $f_i=f_j$.

Obviously, the maximum of the sum in~(\ref{expand}) is achieved if and only if each of
$f_i$ corresponds to a peak of $z_{\rm sf}(f)$. However, we cannot just select the
absolute maximum of $z_{\rm sf}(f)$ at some $f=f^*$ and claim that the global maximum of
$z(\bmath f)$ is at $f_i=f^*$ and is equal to $n\max z_{\rm sf}$. This would imply that
all $f_i$ are equal to each other, which invalidates~(\ref{expand}). Instead of this, we
should locate $n$ \emph{different} tallest peaks of $z_{\rm sf}(f)$ and sum them up to
approximate the global maximum of $z(\bmath f)$.

This approximation allows us to dramatically speed up our Monte Carlo simulations,
although now we can only deal with the cases that are free of spectral leakage. Instead of
scanning the multi-dimensional frequency grid, it is enough to evaluate the
single-frequency periodogram.

\begin{figure}
\includegraphics[width=84mm]{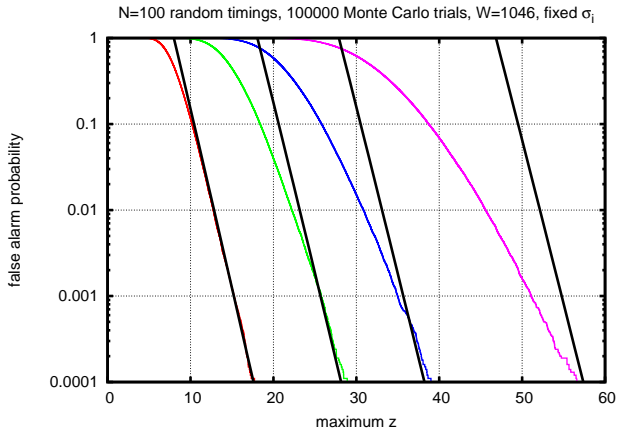}
\caption{The simulated $\FAP$ curves for the multi-frequency periodograms obtained using
the approximation~(\ref{expand}), and the theoretic $\FAP$ estimations~(\ref{fap_mfreq}).
The adopted number of the frequency components was $n=1,2,3$, and $5$ (from left to
right).}
\label{fig_simul_multi}
\end{figure}

The results are shown in Fig.~\ref{fig_simul_multi}. We may notice that the simulated
$\FAP$ curves slightly break the inequality of~(\ref{fap_mfreq}) in the range below the
level of $10^{-3}$. We believe this indicates the inaccuracy of the
expansion~(\ref{expand}) rather then the actual breaking of~(\ref{fap_mfreq}). This is
because in the previous simulation of Fig.~\ref{fig_simul2} the double-frequency $\FAP$
did not exceed the analytic estimation, while in Fig.~\ref{fig_simul_multi} it does.
Taking into account the inaccuracy of~(\ref{expand}) and Monte Carlo uncertainties, our
conclusion is that the analytic estimation~(\ref{fap_mfreq}) does not hide an obvious math
error at least.

\section{Detection sensitivity}
\label{sec_sens}
It is interesting to investigate the detection sensitivity of the multi-frequency
statistic, comparing it with the classic single-frequency one. We will do this in a
simplified framework, in terms of the $\FAP$ estimations~(\ref{fap_dfreq})
and~(\ref{fap_sfreq}). We consider only the case $n=2$. According to the argumentation of
Section~\ref{sec_mot}, we must evaluate in this case three periodograms: two
single-frequency residual periodograms (Foster's SLICK periodograms computed assuming that
one of the periodicities is in the base model) and the only available double-frequency
periodogram.

Let us first assume that amplitudes of the periodic components are equal. Then, in a rough
approximation (discarding the aliasing effects), the maximum values of the
single-frequency periodograms are equal to each other, and according to~(\ref{expand}),
the maximum the double-frequency periodogram is roughly twice the maximum of the
single-frequency ones. This means that we should compare the values of~(\ref{fap_sfreq})
with~(\ref{fap_dfreq}), substituting some $z$ in the first and $2z$ in the second. Notice
that the plain Lomb-Scargle periodogram of the raw data is equal, in this approximation,
to the sum of the individual single-frequency periodograms mentioned above. Therefore, it
will posess two distinct peaks at their relevant frequencies, both having approximately
the same height of $z$.

\begin{figure}
\includegraphics[width=\linewidth]{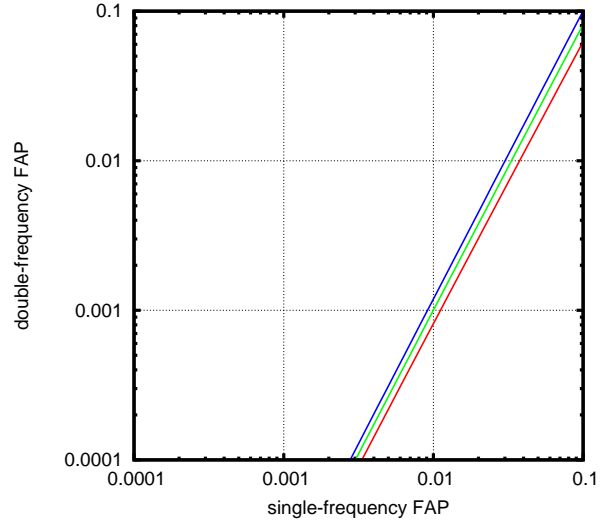}
\caption{Comparison of the detection efficency of the single- and double- frequency
periodograms in terms of the analytic $\FAP$ estimations from~(\ref{fap_sfreq})
and~(\ref{fap_dfreq}). The three plotted curves correspond to $W=100$, $1000$, and
$10000$. The actual signal is assumed to involve two periodicities of the same amplitude,
so the maximum value of the single-frequency periodogram is roughly half of that of the
double-frequency one.}
\label{fig_FAP}
\end{figure}

We plot the relevant $\FAP$ comparison graphs in Fig.~\ref{fig_FAP}, for $W=100$, $1000$,
and $10000$. We may notice two things: the double-frequency periodogram appears in this
case definitely superior over the single-frequency one in terms of the sensitivity, and
this advantage is almost independent of $W$. This means that when our periodic components
have equal amplitudes, it is unlikely that the double-frequency periodogram may disprove
the single-frequency detections.

\begin{figure}
\includegraphics[width=\linewidth]{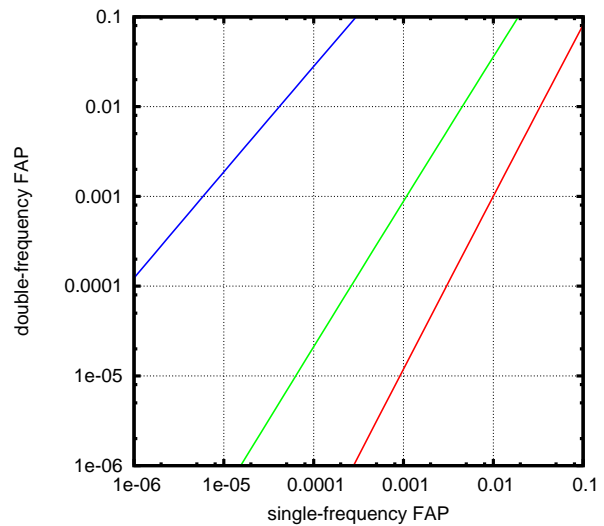}
\caption{Similar to Fig.~\ref{fig_FAP}, but now we fix $W=1000$ and vary the amplitude
ratio of the sinusoidal components: $1\!:\!1$, $1.2\!:\!1$, and $2\!:\!1$ (curves from
right to left).}
\label{fig_FAP2}
\end{figure}

When we consider the cases of unequal amplitude ratio (Fig.~\ref{fig_FAP2}), we can find
that it affects crucially the sensitivity of the double-frequency periodogram relatively
to the single-frequency one. For the amplitude ratio $1.2$ the sensitivities become
roughly similar, and after that the double-frequency $\FAP$ becomes much larger than the
single-frequency one. Therefore, in the more frequent cases with unequal amplitudes, the
extra statistical verification by the double-frequency periodogram is mandatory, since it
may easily disprove our single-frequency detections.

However, we may notice that in terms of the detection thresholds (critical $z$ levels) the
difference between the single- and double-frequency periodograms is not that huge as it
may seem when we compare $\FAP$s in Fig.~\ref{fig_FAP2}. From~(\ref{fap_sfreq})
and~(\ref{fap_dfreq}) we conclude that the difference between the $z$ levels is roughly
the logarithm of the relevant $\FAP$ ratio.

\section{A double-frequency example}
\label{sec_examples}
In the above sections, the reader may become convinced that we preach a single purpose of
the multi-frequency periodogram: disappoint a hasty person who claimed a detection of
several periods too soon. This not the only purpose, of course. Let us provide some
demonstration of how the double-frequency periodogram may work in a constructive rather
than destructive manner. We consider the following model example. The $N=150$ data points
are clumped in $5$ groups (each contains $30$ randomly distributed points) separated by
gaps. The gaps cover $43\%$ of each such ``data+gap'' chunk. The data are noisless and
contain only a signal, which is a sum of two cosinusoidal components with $f_1=0.9$~Hz and
$f_2=1.1$~Hz and equal amplitudes. The phases of the components are $0^\circ$ and
$145^\circ$.

\begin{figure*}
\includegraphics[width=0.8\linewidth]{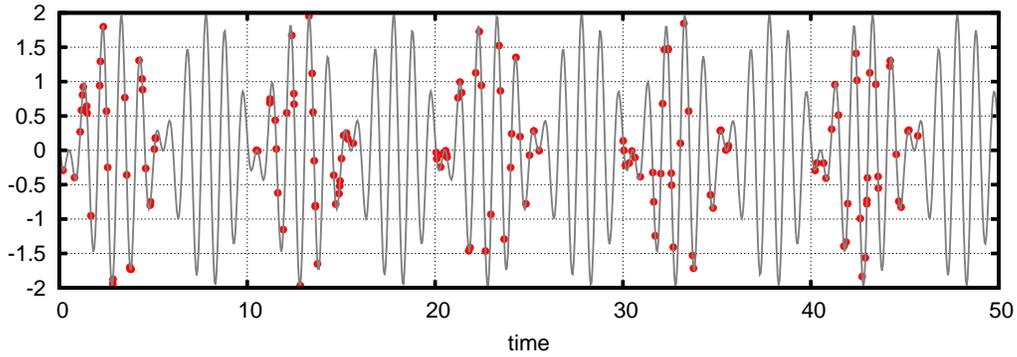}
\caption{The signal and data of the model example of Section~\ref{sec_examples}. The
signal contains two periodic components which generate the beating, and the dataset is
such that only each odd beating cycle is sampled (see text for the detailed description).
The noise is absent.}
\label{fig_beating}
\end{figure*}

\begin{figure*}
\includegraphics[width=0.49\linewidth]{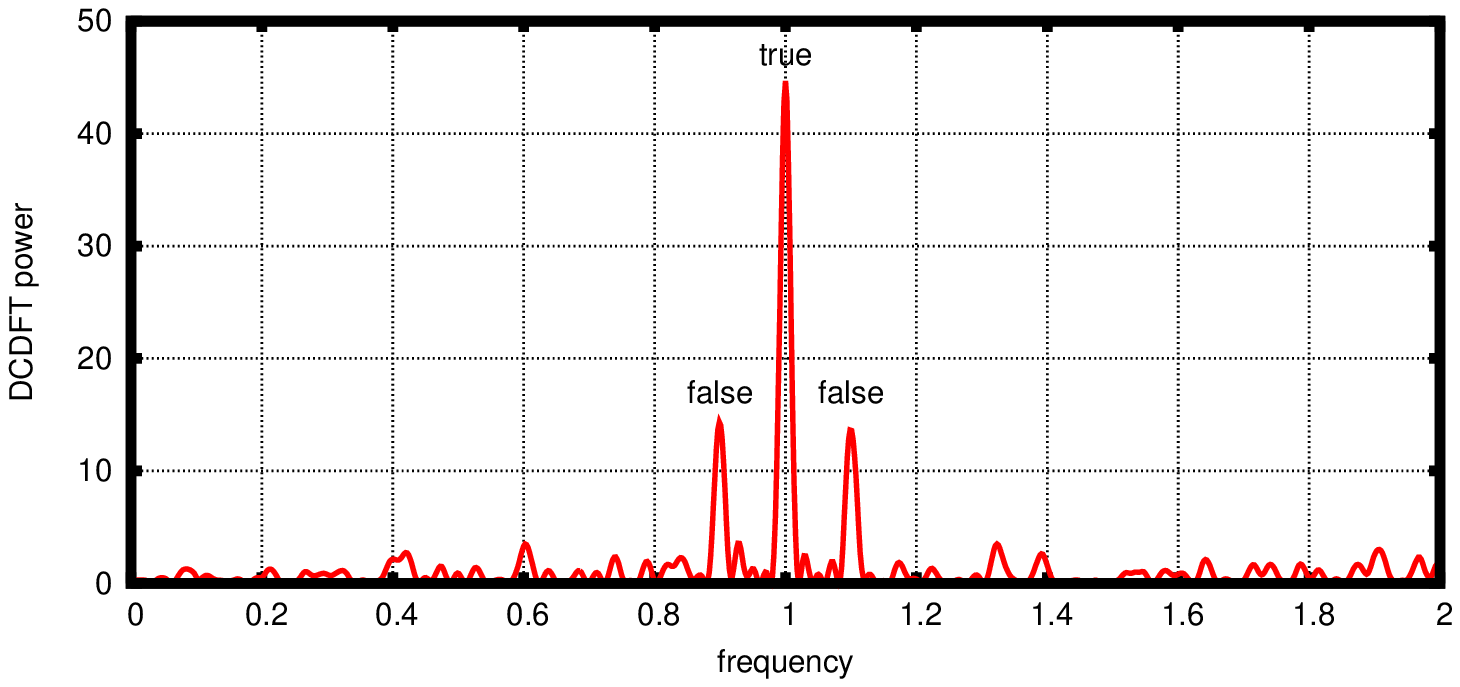}
\includegraphics[width=0.49\linewidth]{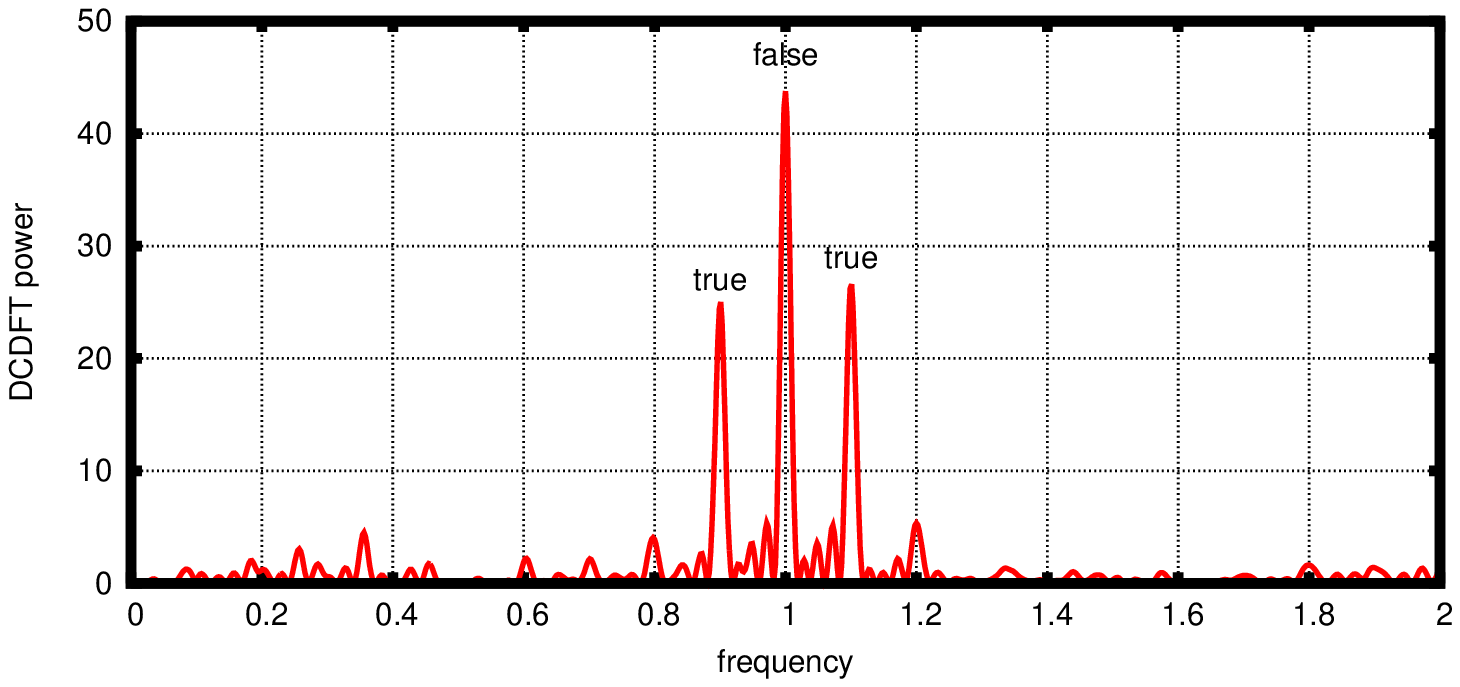}
\caption{The Lomb-Scargle periodograms of two synthetic datasets: of the one shown in
Fig.~\ref{fig_sf} with two periodicities at $0.9$~Hz and $1.1$~Hz (right graph), and of a
similar one containing only a single periodicity at $1.0$~Hz (left graph). The time series
involves periodic gaps generating an aliasing frequency of $0.1$~Hz (see text for the
detailed description). The noise is absent.}
\label{fig_sf}
\end{figure*}

This signal represents a beating process shown in Fig.~\ref{fig_beating}. We can see that
the gaps in the data are such that only each odd beating cycle is sampled. The DCDFT
periodogram of these data shows no much difference with the case of a single sinusoid at
$f=1.0$~Hz (Fig.~\ref{fig_sf}). In practice we would be unable to distinguish such single-
and double-frequency case. The only difference is that the side peaks in the
double-frequency case are larger than for the single-frequency one. However, in practice
this would just make us to think that the spectral leakage is a bit stronger than it
actually is. Therefore, the we are unable to find a correct model for the signal in
Fig.~\ref{fig_beating}: the single-frequency periodogram would direct us to a wrong way
from the very beginning stage of the analysis.

\begin{figure*}
\includegraphics[width=0.49\linewidth]{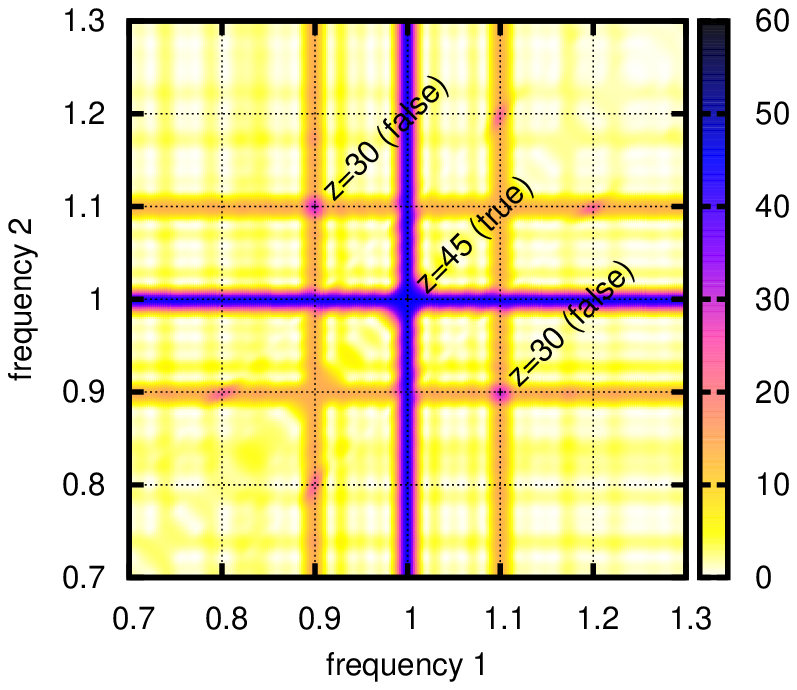}
\includegraphics[width=0.49\linewidth]{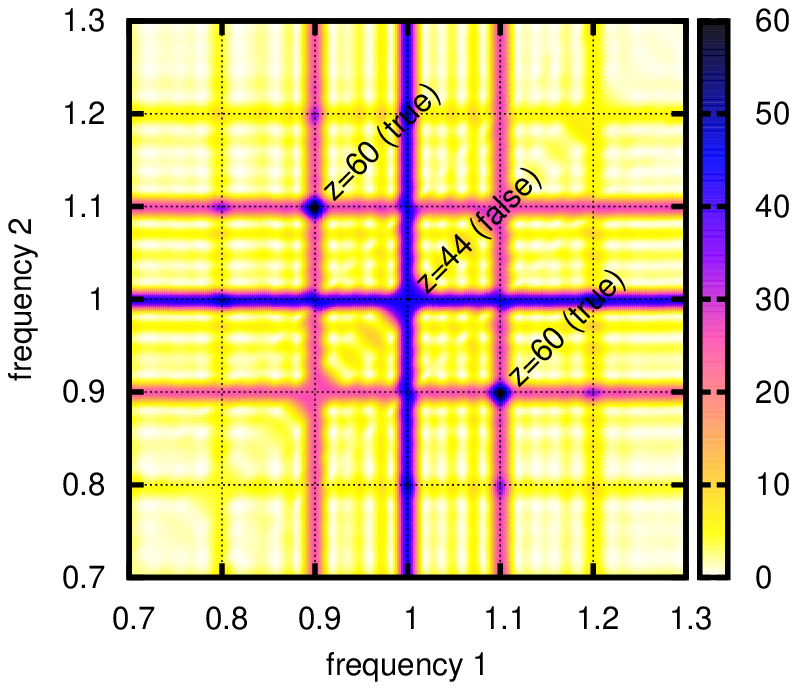}
\caption{The double-frequency periodograms plotted for the same datasets as
in Fig.~\ref{fig_sf}. We label the highest peaks with the corresponding maximum
periodogram values.}
\label{fig_df}
\end{figure*}

However, armed with the double-frequency periodogram, we can find the correct solution of
the problem immediately. As we can see in Fig.~\ref{fig_df}, this periodogram reveals the
correct period pair for the double-frequency signal, and simultaneously it does not
generate any undesired additional periods for the single-period data.

Of course, in practice the success of the analysis would also depend on the signal/noise
ratio, but the Lomb-Scargle periodogram failed already for entirely noiseless data that we
have just considered.

Our general conclusion is that the double- and multi-frequency periodograms may appear
rather useful in certain especially difficult time-series analysis tasks. They are able to
reveal correct periodic solutions in the cases when single-frequency periodograms fail.

\section{Conclusions}
On itself, it is usually quite easy to invent a sophisticated periodogram to satisfy the
demands of some specific data-analysis task. For example, the multi-frequency periodogram
that we considered here was known for almost two decades already. One of the principal
obstacles that put a strict limitation on the practical use of such periodograms is the
need of a simultaneously rigorous, general, and computationally efficient approach to
evaluate the significance levels associated with these new periodograms. Even for the
classic Lomb-Scargle periodogram the evaluation of these significance levels represented a
substantial difficulty over decades.

We believe that the $\FAP$ estimation approach based on the generalized Rice method, that
we are using extensively during last 5 years, satisfies all these requests. It inherits
the generality and rigorous basis of the Rice method. Also, it often leads to entirely
analytic and self-closed results that work according to a simple principle ``just
substitute''.

The importance of the detection significance levels for the multi-frequency periodogram is
even further emphasized, because it is not just some fancy multi-frequency periodogram
that we may use or may refuse to use. As we have discussed, the need to assess the
multi-frequency $\FAP$ still persists even when we detect periodicities in a sequential
single-frequency manner. If we wish to have nothing common with any multi-frequency
periodograms, we may produce an increased number of false detections.

We would like draw some more attention to the comparison of the Lomb-Scargle $\FAP$
formulae~(\ref{fap_sfreq}) with e.g. its double-frequency analog~(\ref{fap_dfreq}).
Remarkably, they looks rather simialar to each other. In fact, we could easily guess the
correct powers of $W$ and of $z$ in~(\ref{fap_dfreq}), based on~(\ref{fap_sfreq}), even
without any sophisticated calculations, just by taking into account the increase of the
number of the free parameters of the signal. However, it would be impossible to guess the
non-trivial and important coefficient of $\pi/16\approx 1/5$. It could be derived only by
means of a rigorous application of the generalized Rice method, as we have done here.

So far in the paper, we paid little attention to the task of practical computation and
maximization of the multi-frequency periodogram. Obviously, this may constitute a
challenge already for $n\geq 3$. In the next Paper II, an efficient parallellized
computation algorithm is to be presented. The beta version of this algorithm is already
available for download at \texttt{http://sourceforge.net/projects/fredec/}, although with
only a little documentation until Paper II.

\section*{Acknowledgments}
This work was supported by the Russian Foundation for Basic Research (project No.
12-02-31119 mol\_a) and by the programme of the Presidium of Russian Academy of Sciences
``Non-stationary phenomena in the Objects of Universe''. I would like to express my
gratitude to the anonymous reviewer for providing constructive suggestions.

\bibliography{multifreq}
\bibliographystyle{mn2e}

\bsp

\label{lastpage}

\end{document}